# Recent advances in pulsed-laser deposition of complex-oxides

H. M. Christen and G. Eres

Materials Science and Technology Division, Oak Ridge National Laboratory, Oak Ridge, TN 37831, USA

**Abstract**

Pulsed-laser deposition (PLD) is one of the most promising techniques for the formation of complex-oxide heterostructures, superlattices, and well-controlled interfaces. The first part of this paper presents a review of several useful modifications of the process, including methods inspired by combinatorial approaches. We then discuss detailed growth kinetics results, which illustrate that 'true' layer-by-layer (LBL) growth can only be approached, but not fully met, even though many characterization techniques reveal interfaces with unexpected sharpness. Time-resolved surface x-ray diffraction measurements show that crystallization and the majority of interlayer mass transport occur on time scales that are comparable to those of the plume/substrate interaction, providing direct experimental evidence that a growth regime exists in which non-thermal processes dominate PLD. This understanding shows how kinetic growth manipulation can bring PLD closer to ideal LBL than any other growth method available today.

## 1. Introduction

Interfaces and crystalline superlattice materials are of increasing interest to a large and growing fraction of the condensed matter physics community. In fact, the deterministic synthesis of such completely artificial crystalline structures allows us to go beyond equilibrium materials in exploring new properties, developing new functionalities, and analyzing fundamental physical processes.

Transition-metal oxides are particularly interesting building blocks for such structures, as they possess a great number of interesting intrinsic properties [1]. However, the precise assembly of such layers into artificial superlattices requires atomic-scale control. Pulsed-laser deposition [2-5], long known as the tool of choice for the growth of complex-oxide materials, has recently been applied to the growth of interfaces [6-7] with a sharpness that was previously thought to be obtainable in molecular-beam epitaxy methods [8,9] but not PLD. With the prospect of forming such artificial materials, it has become critically important to understand the fundamental limits in obtaining atomically flat growth surfaces. This is complicated by the scarcity of tools to investigate—at an atomic scale—the quality of an interface: *final* surfaces can be characterized, for example, by atomic-force microscopy (AFM) (with a lateral spatial resolution of many unit cells), but *embedded* interfaces can potentially be much different. Scanning transmission electron microscopy (STEM) gives access to a projection of a specimen with thickness of a few tens of nanometers. Broadening of the observed interfaces is often attributed to various types of defects and surface steps as well as to an intrinsic dechanneling of the electron beam as it traverses the sample. Thus, it is rarely possible to distinguish between a true atomically-flat interface and one with a non-vanishing roughness of—for example—a single unit cell.

Surface x-ray diffraction (SXRD) provides an alternative means to explore the question of how smooth a layer can be ultimately for a given growth environment, and the results illustrate the importance of considering the fundamental limits beyond the current microscopic techniques' capabilities. True layer-by-layer growth is known to be fundamentally impossible by any currently available growth technique [10], and our time-resolved SXRD observations clearly confirm this observation for homoepitaxial PLD of $SrTiO_3$. In addition, however, our results provide an explanation for the surprising success of PLD and provide guidance for further improvements of the method.

The purpose of this paper is to discuss the fundamentals of growth in the context of SXRD observations of the PLD process. Before that, however, we give a brief overview of the pulsed-laser deposition method as applied to metal oxides, without pretending to provide a comprehensive review. Such reviews can be found in the literature, where the growth of complex metal-oxide films by PLD is compared to results obtained by other techniques [5,11]. In this paper, however, we aim to provide a tutorial-style introduction to PLD, focus on specific issues that need to be addressed, and describe modifications of the process that have allowed us to apply the method to a number of systematic investigations. This paper first describes efficient techniques (compositional-spread, temperature-gradient, etc.), and then focuses on the details of nucleation and growth, with a strong emphasis of results from time-resolved SXRD.

## 2. Basic concepts

2.1 Development of PLD

The use of a pulsed laser to induce the stoichiometric transfer of a material from a solid source to a substrate, simulating earlier flash evaporation methods, is reported in the literature as early as 1965 [12], where films of semiconductors and dielectrics were grown using a ruby laser. Pulsed-laser evaporation for film growth from powders of $SrTiO_3$ and $BaTiO_3$ was achieved in 1969 [13]. Six years later, stoichiometric intermetallic materials (including $Ni_3Mn$ and low-$T_c$ superconducting films of $ReBe_{22}$) were produced using a pulsed laser beam [14]. In 1983, Zaitsev-Zotov and co-workers reported for the first time superconductivity in pulsed-laser evaporated $BaPb_{1-x}Bi_xO_3$ films after heat-treatment [15]. The real breakthrough for PLD, however, was its successful application to the in-situ growth of epitaxial high-temperature superconductor films in 1987 at Bell Communications Research [16].

Since then, PLD has been used extensively in the growth of those high-temperature cuprates and numerous other complex oxides, including materials that cannot be obtained via an equilibrium route. Early on, it has been shown that the processes in the growth of materials from a PLD plume are fundamentally different than those found in thermal evaporation [17]. The method has been successful for the film synthesis of Y-type magnetoplumbite (with a c-axis lattice parameter of 43.5Å) [18] and garnets with 160 atoms per unit cell [19]. As the PLD process became better controlled and more sophisticated, the term "laser-MBE" was introduced to describe a PLD system in which layer-by-layer growth is achieved and monitored by RHEED (reflection high energy electron diffraction), or simply for PLD in ultra-high vacuum (UHV). This terminology, of course, is somewhat inaccurate, as a laser plume always contains a combination of ions, electrons, and neutral particles and is thus not a molecular beam. Nevertheless, "laser-MBE" has been used successfully to go beyond the codeposition of all components of a complex oxide by instead to depositing single layers of SrO and BaO sequentially [20] and intercalating SrO layers in manganites [21]. However, the term is often used even when ablation occurs from a complex target [22-24], at which point "laser-MBE" simply implies "UHV-PLD" or "*in situ* monitored PLD."

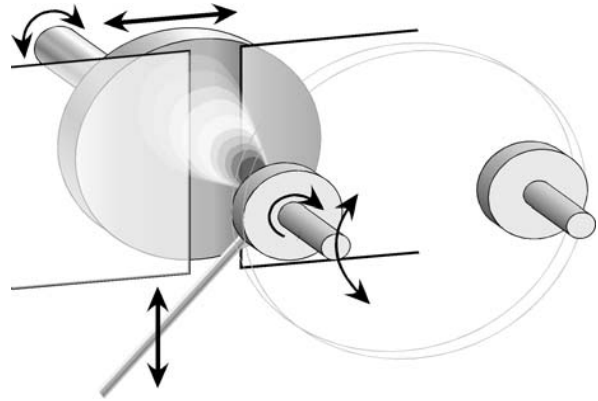

**Figure 1.** Schematic representation of the main components in a standard PLD process. Arrows in the figure represent axes of motion that can be controlled and synchronized. Two targets are shown in the foreground, and the laser beam (entering from the lower left) impinges onto one of them, forming a plasma plume. This plume expands towards the substrate heater, shown behind the slit-shaped aperture. The aperture is not typically present but can be used for compositional-spread approaches as described in the text. Reprinted with permission from Ref. 79, Copyright 2005, IOP Publishing Ltd.

The technique of PLD is conceptually simple as illustrated schematically in Fig. 1. A pulsed laser beam leads to a rapid removal of material from a solid target and to the formation of an energetic plasma plume, which then condenses onto a substrate. In contrast to the simplicity of the technique, the mechanisms in PLD – including ablation, plasma formation, plume propagation, as well as nucleation and growth – are rather complex.

In the drawing of Fig. 1, the target is assumed to be a ceramic disc, with ablation occurring on the flat surface. Other arrangements (shape of targets, positioning of substrates) are possible. For example, the substrate can be placed elsewhere with respect to the plume, and can be positioned such as to have its surface parallel to the direction of the plume propagation rather than perpendicular [25] or even in the plane of the target [26]. Other arrangements are discussed in earlier reviews [for example, Ref. 5]. It is one of the goals of this paper to show how modifications of the basic arrangement of Fig. 1 can result in very useful approaches for the rapid exploration of new materials.

2.2 Ablation and plasma formation

In the process of laser ablation, the photons are converted first into electronic excitations and then into thermal, chemical, and mechanical energy [27,28], resulting in the rapid removal of material from a surface. This process has been studied extensively because of its importance in laser machining. Heating rates as high as $10^{11}$ K/s and instantaneous gas pressures of 10 – 500 atm. are observed at the target surface [29]. The laser-solid interaction mechanisms may depend on the laser wavelength; in fact, significant changes in the energetics of species in a plume resulting from ablation of carbon using KrF (248nm) and ArF (193 nm) excimer lasers are observed [30], having a large effect on the growth of diamond-like carbon films. The most important effect of the laser's wavelength is its determination of the penetration depth. Most of the energy should be absorbed in a very shallow layer near the surface of the target to avoid subsurface boiling which can lead to a large number of particulates at the film surface. However, the absorption of photons by oxygen molecules and optical elements in the beam path determines a lower practical wavelength limit of approximately 200 nm.

For relatively long pulse durations, such as the tens of nanoseconds typical for excimer lasers, there is a strong interaction between the forming plume and the incident beam, leading to a further heating of the species. This may explain experiments of $YBa_2Cu_3O_{7-\delta}$ film growth where, for a given laser energy density at the target surface, ablation using a KrF excimer laser (248 nm, ≈ 25 ns pulse duration) resulted in far superior films than ablation using a frequency-quadrupled Nd:YAG (266 nm, ≈ 5 ns pulse duration) [31]. Similarly, certain aspects of a dual-laser approach [32], where a $CO_2$ laser pulse with a 500 ns duration is allowed to interact with the plume formed by the ablation using a KrF excimer laser, have been attributed to increased laser heating of the plasma.

Comparing PLD to pulsed-electron deposition(PED) reveals several intersing aspects of the deposition process. In PED [33-36], an electric discharge rather than a laser pulse creates a plasma, and the energy density (integrated over the pulse duration) at the target surface is very similar to that obtained in PLD. However, the pulse duration (100 ns) is significantly longer. Compared to PLD, the PED process shows more significant deviations from stoichiometry, with strong variations as a function of the position on the substrate [37]. This indicates the importance of a very dense plasma near the target surface in order to create a plume in which all species – regardless of their mass – expand with an identical angular distribution.

Finally, the laser fluence at the surface of the target has to exceed a certain threshold, which in many configurations ranges from 1 – 3 $J/cm^2$ for a 25 ns pulse. A quite different value of 0.3 $J/cm^2$ has been found to be optimal for the ablation of $SrTiO_3$ from a single crystal (rather than ceramic) target in a background of $10^{-6}$ Torr (rather than the typical 5 – 500 mTorr) and using a laser with a comparatively fast rise time [38]. Even so, the required energies per pulse are fairly high and most readily achieved with excimer lasers [39]. Lasers using KrF excimers (248 nm, typically 20 – 35 ns pulse duration) have been used most often in PLD, but successful film growth has also been achieved using ArF (193 nm) [40-42] and XeCl (308 nm) [43-47] excimers.

Many "ultrafast lasers" deliver less energy per pulse, but with a much shorter pulse duration (thus high instantaneous power) and a higher repetition rate than excimer lasers. For chemically less complex materials such as simple oxides (where stoichiometry is not an issue and thus the plasma density may not matter as much) film growth has indeed been possible using a variety of lasers, including hybrid dye/excimer lasers (248 nm, 500 fs) [48] and femtosecond Ti-sapphire lasers [49-51]. A 76 MHz, 60 ps mode-locked Nd:YAG laser has been used successfully for the "ultrafast ablation" and growth of amorphous carbon [52]. More recently, similar techniques have been applied to the growth of more complex materials, such as $Ge_{33}As_{12}Se_{55}$ [53], leading to optimism for the use of these different lasers.

2.3 Plume propagation

Plume propagation has been studied extensively using optical absorption and emission spectroscopy combined with ion probe measurements [29,54,55], and does not need to be discussed in detail here. Neutral atoms, ions, and electrons travel at different velocities, and strong interactions between the species of the plasma and the background gas are observed. In fact, it is sometimes assumed that some degree of thermalization needs to occur in order to obtain good film growth and to avoid resputtering of the growing film by the most energetic ions in the plume [56]. Assuming that most of the species in the plume should be fully thermalized at precisely the time they reach the substrate (i.e., having equal lateral and forward velocities), a simple model predicts that the optimal growth rate should be close to 1 Å per pulse [57,58]. This is rather close to the actually observed values for many experiments where stabilization of a complex material is the main goal. However, the precise formation of superlattice materials, especially comprised of $SrTiO_3$ and related perovskites, is often best achieved at much lower deposition rates (requiring hundreds of laser pulses per unit cell) [6].

2.4 Control of stoichiometry

The stoichiometric removal of material from a solid target is undoubtedly the single most important factor in the success of PLD. For a vast majority of ceramic targets, and for ablation rates that result in a dense plasma as described above, the removal of material does indeed preserve stoichiometry. This is particularly true after an initial "preablation" process, i.e., the exposure of the target surface to the laser irradiation for some time before deposition in order to obtain a steady-state (which, if one of the elements is more volatile than others, results in an enrichment of the target surface of the less-volatile component).

Stoichiometric removal of the material from the target, however, does not necessarily translate into the growth of stoichiometric materials, as not all elements get incorporated at the same rate (often referred to as a "sticking coefficient"), some re-sputtering can occur [56], and volatile elements may re-evaporate from the growth surface. When growing materials containing an element that is considerably more volatile than others, such as $KNbO_3$ (with potassium being the more volatile species), the use of an additional source is often required. This can be done, for example, by using a rotating segmented target, consisting of $KNbO_3$ and $KNO_3$ – the latter being an additional potassium supply [59,60]. Similarly, non-stoichiometric targets are often used to compensate for the loss of Bi or Pb. In all of these cases, it is generally observed that it is possible to work in a regime of significant excess of the volatile component, all of which re-evaporates beyond the amount needed to form the stoichiometric compound.

For oxide materials, proper control of the oxygen content is of paramount importance. The fact that PLD is possible in a broad range of background pressures aids especially in the formation of ferroelectric materials, for which no other method has produced better properties than those achievable by PLD [61,62].

## 3. Implementations of PLD

In addition to the stoichiometric removal of material, one of the great advantages of PLD is the intrinsic flexibility of the approach, which results from the use of solid starting materials held at room temperature, and an energy source external to the vacuum system. Mechanical actuation of targets and substrate is therefore relatively simple and can be synchronized with the laser firing. The pulsed nature of the optically delivered energy makes it possible to utilize mechanical motion in a way that leads to precisely controlled alloys, composition gradients, and superlattices, as we describe in this section.

3.1 Alloy formation by sequential ablation from multiple targets

The fact that each laser pulse (under the appropriate conditions) results in the deposition of far less than a monolayer of material can be utilized to "mix" materials from separate sources. A simple way to implement this is illustrated in Fig. 2a for the growth of ferroelectric $KTa_{1-x}Nb_xO_3$ films [63]. A target consisting of three segments is rotated around a point along the interface between the $KNbO_3$ and the $KNO_3$ portions, off-set from the center of the assembly by a distance $d$. The laser impinges on a circular track with radius *R*. Two goals are simultaneously achieved by this method: First, the volatility of potassium is compensated for by the additional $KNO_3$ segment [59], as mentioned above. Second, ablation occurs sequentially from the $KTaO_3$ and $KNbO_3$ portions of the target. If the laser repetition rate and the target rotation velocity are chosen appropriately for example, such that the laser repetition rate is not a multiple of the target rotation rate), the resulting film has a composition of $x = \frac{1}{\pi}\left[\alpha + \arcsin\left(\frac{d}{R}\sin\alpha\right)\right]$ (as shown for different angles α in Fig. 2b). Here we make the assumption of equal ablation rates for both constituents, but the approach can readily be generalized. Obviously, this method can be simplified for the case of two materials rather than the case presented here with the additional $KNO_3$ component.

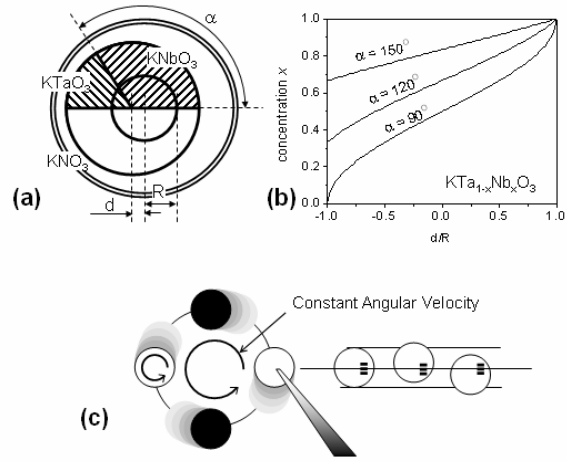

**Figure 2.** Alloy formation by the sequential deposition of sub-monolayer amounts of each constituent. (a) Schematic representation of the segmented target, mounted off-center from the axis of rotation by a distance *d*, with the laser impinging at a distance *R* from the center, and shown for the case of $KTa_{1-x}Nb_xO_3$, where an additional $KNO_3$ segment is used to compensate for the volatility of K. The concentration can be varied simply by adjusting the ratio of *d/R*, as shown in (b). (c) Alternative approach for alloy formation based on the rapid exchange of multiple targets, with the firing of laser pulse bursts being synchronized with the target position. Different time delays can be used to achieve the same effect as target rastering in a stationary-target method.

Other approaches have been introduced to obtain two-target mixing for alloy formation. For example, a rod-shaped target, formed of two sections, can be translated along its rotation axis to expose the different components to the laser beam [64]. Similarly, the common geometry of a multi-target carousel can be used for such alloy formation simply by exchanging the target before a full monolayer of the material has formed [65]. This may result in reduced overall deposition rates if the target exchange mechanism is slow. In our laboratory, we have overcome this limitation by imposing a continuous rotation of the target carousel, as shown in Fig. 2c, combined with a laser triggering mechanism based on the target position: bursts of laser pulses are fired at precisely the moment when the target is in the correct position. With this, the target carousel rotation can be as fast as 1 revolution per second, and the effective (average) laser repetition rate is 2 – 10 Hz (with a 10 ms delay between successive laser pulses fired onto an individual target). This method has been applied to various materials, such as the ferroelectric $Bi(Fe_{1-x}Cr_x)O_3$ alloys [66], and its use in the formation of vertical composition gradients (i.e., in the direction of film growth) is a straight-forward extension.

3.2 Continuous compositional spread

As mixing between targets results in the formation of alloy materials, it is a natural extension of this method to combine the mixing with a lateral translation of the substrate in order to obtain spatial variations of the composition on a larger substrate. Such an approach is motivated by the observation that combinatorial materials science has enjoyed great success in the chemical and pharmaceutical industries, with successes in thin-film research still being comparatively rare. This is in part due to the great technical challenge involved in the characterization of the resulting materials. An approach for the parallel multi-sample synthesis is therefore needed that yields samples large enough for conventional measurement techniques to be used, even if a smaller number of materials can be simultaneously synthesized and studied.

Earlier implementations of combinatorial PLD approaches [67-70] were based on the room-temperature deposition of a precursor material, which is then converted to a complex perovskite in a post-deposition annealing process. The advantage of these discrete combinatorial methods – which are based on the precise positioning of delicate masks – is the large number of compositions that can be synthesized, at the expense, however, of a synthesis method that is fundamentally different from the most successful implementations of PLD. Later versions of PLD-based discrete combinatorial methods have been used for the formation of complex materials directly at elevated temperatures [71-73]. Unfortunately, even for the relatively small number of compositions explored simultaneously, the small substrate size still leaves the requirement of specialized characterization techniques.

One approach to obtain a laterally varying composition across a large substrate is to use the naturally observed spatial growth rate variations in PLD, similar to what has been done early in sputter-based compositional-spread methods [74,75]. This is in fact possible using PLD with synchronized substrate motion and laser firing [76]. However, in addition to the growth rate, the energetics of the deposited species in PLD may also vary as a function of position on the substrate, and the method thus suffers from the simultaneous variation of two parameters across the sample surface.

This difficulty can be overcome by inserting an aperture between the target and the substrate, as shown in Fig. 1. The combined translation of the substrate behind this aperture and the exchange of targets after deposition of less than a monolayer results in linear composition gradients. As we have shown in Ref. [77], the position of laser "trigger points" can be calculated such that a linear composition variation is obtained simply by firing the laser each time one of the trigger points is aligned with the center of the aperture. This produces a very stable and easy-to-implement approach, sending a trigger signal to the laser each time the heater position (or an encoder on the motor that drives the heater translation) coincides with one of these pre-calculated trigger points. The oscillatory heater motion can then be chosen such as to optimize its speed: in our configuration with a travel distance of ~ 50 mm, each pass requires approximately 1 s (requiring in most cases effective maximum laser repetition rates between 50 and 100 Hz). This method has been applied successfully to the synthesis of complex transition-metal oxides, such as alloys between paramagnetic $CaRuO_3$ and ferromagnetic $SrRuO_3$ [77].

3.3 Controlling the lateral thickness variation

Just as repeated back-and-forth passes of the substrate behind the aperture in Fig. 1 can be used to repeatedly deposit sub-monolayer amounts in order to form alloys, the method is readily adapted to situations where simple thickness gradients are desired. Such samples are important, for example, in the study of thickness effects on the properties of films. The method can also be used for entirely different applications: orthogonally overlapping wedges of metallic films can be used, for example, in the study of catalysis of carbon nanotubes [78].

Obtaining good thickness uniformity – rather than controlled thickness gradients – is often important, but impossible by simply depositing onto a stationary substrate. Using again the approach of synchronized laser firing and substrate positioning, the uniformity can be controlled [79] either on larger wafers or on a row of samples as necessary for the temperature-gradient method described below.

3.4 Temperature-gradient approaches

The growth temperature is often the most critical parameter in any film deposition experiment. Determining the correct substrate temperature is therefore a necessary but time-consuming first step in the exploration of new materials. In the spirit of the above-described "multi-sample" approaches, it is possible to deposit simultaneously onto multiple samples, each held at a different temperature. Such an approach was first reported more than 40 years ago [80], but has only recently been applied to complex oxides [81,82]. In these recent approaches, a large temperature difference (~300 °C) is obtained across a 10mm sample by the use of laser heating. Again realizing the importance of traditional characterization techniques, however, we have implemented an approach in which the temperature gradient is much smaller (a change of 600 °C across a distance of 70 mm) [83]. Using samples that are typically 2 mm wide in the direction of the gradient and 10 mm long in the orthogonal direction, the temperature difference across each sample is about 15 °C. The approach thus provides sufficient temperature uniformity for initial studies, and yields samples that are easily characterized using traditional x-ray diffraction or transport methods.

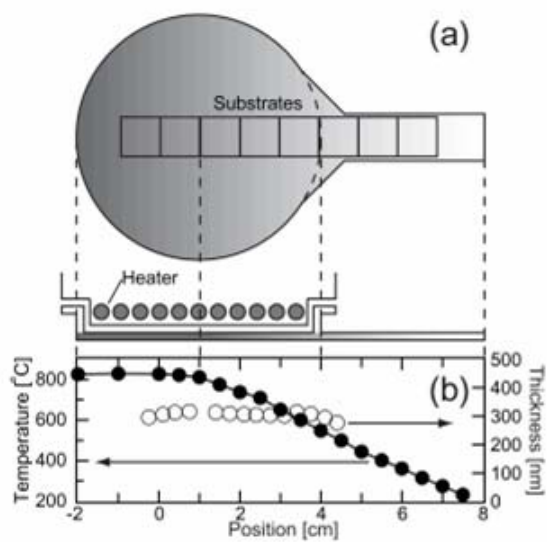

**Figure 3.** (a) Schematic representation of the substrate heater with a metallic plate yielding a smoothly varying temperature profile. Additional heat shields are added and temperature-isolating slots are cut into this plate (not shown). (b) Temperature variation across the substrate plate (filled circles, left scale) and obtained film thickness (open circles, right scale), showing satisfactory temperature linearity and thickness uniformity. Reprinted with permission from Ref. 83, Copyright 2004, American Institute of Physics.

Fig. 3 shows a schematic of the method. A radiative heater (based on a Pt-alloy heat element) is used to heat a metallic substrate holder, which is cut into a shape such that a protruding "finger" results on local heat loss. Combined with additional shields and slots cut into this plate (not shown), a linear temperature variation, as shown in Fig. 3b, is obtained. The method has been applied to the study of electro-optic materials [83] and the determination of crystallization temperatures of a series of rare-earth scandates as candidate high-k dielectric gate materials [84].

3.5 Superlattice growth

In section 3.1 above, we have shown that sequential ablation from separate targets can lead to alloy formation if much less than one unit cell is deposited in each cycle. The approach to form heterostructures and superlattices is conceptually similar, except that at least one "complete layer" is grown before depositing the next material. Obviously, in order to form a true superlattice, the roughness of each interface must be significantly smaller than the thickness of these layers. As we show below in the discussion of growth kinetics, this is a non-trivial requirement when the layer thickness decreases to one unit cell. We use our SXRD data to show that current understanding of nucleation and growth predicts that no growth method is capable of producing a completely full layer that is exactly one unit cell thick with true atomic-scale sharpness. The only possible exception is step flow growth, for which in-situ monitoring techniques fail.

Despite these fundamental considerations, the ease by which PLD allows us to alternate between different materials has been used by numerous groups to apply the technique to superlattice growth. The simplest approach is to calibrate the growth rates for each material (in terms of the amount of material deposited per laser pulse) and then grow the superlattice by counting laser pulses [60,85-88]. The use of RHEED makes it possible to track the number of layers deposited in real-time, as the intensity oscillates with a periodicity equal to the time required for the deposition of a monolayer [89].

Superlattice peaks in x-ray diffraction scans are a clear indication that a periodic structure has been obtained; however, without actual calculation (numerical modeling) of the peak intensities, these data only demonstrate periodicity, not interface sharpness. Z-contrast scanning transmission electron microscopy (STEM) has become a widely accepted and frequently-used tool to analyze interfaces, by visualizing a projection of a specimen with a thickness of a few tens of nanometers. Here, interfaces that appear atomically sharp within the limits of the technique [6,90] have been observed. In fact, the quality of PLD-grown samples has now reached a level that was previously thought to be achievable exclusively by MBE. As we mentioned above, steps in the substrate surface and the natural and unavoidable broadening of the electron beam as it traverses the sample (dechanneling) currently make it impossible to distinguish between a perfect atomically-sharp interface and a partially diffuse layer of a single unit cell thickness (most data presented in the literature show a width—apparent or real—of more than one unit cell). Surface x-ray diffraction is currently the only technique that can identify and probe intermixing on a single unit cell level [91].

In order to understand the fundamental limits to atomically sharp interfaces formation in current methods, and to determine how to modify these methods to yield the desired perfectly-flat layers, we next turn our attention to basic considerations of nucleation, island growth, coalescence, and layer filling, before discussing the new understanding gained from time-resolved SXRD studies.

## 4. Simple models of nucleation and growth

4.1 Step flow and island formation

Most mechanisms in film growth are strongly materials-dependent – the formation of an epitaxial perovskite film involves a different mechanism than the deposition of a gold layer. For the purposes of this paper, we are concerned primarily with the synthesis of epitaxial perovskite films at the early stages of growth. Even though the precise mechanisms leading to the crystallization of these complex materials are still unknown, it is illustrative to begin this section with a few simple and generic considerations. A more detailed description of various growth models is found in [92]

The simplest possible model of film growth is that of atoms landing on a surface, where they randomly select a site at which they remain immobile. If the probabilities of "sticking" are equal for all sites (including sites with nearest-neighbors, and those on top of already-deposited species), then the deposition of $N$ particles onto $N$ sites (i.e., one monolayer) leaves 37% (=1/$e$) of the surface uncovered. Note that real systems are quite different, and even an ideal ball model on a hexagonal lattice is fundamentally different, as there are no sites above a single adatoms. At any finite temperature, the picture of immobile balls changes: the deposited species ("adatoms") have a non-vanishing diffusivity $D$ and thus remain mobile at the surface until they are immobilized when they encounter an energetically favorable site. In the absence of extrinsic nucleation sites (e.g. defects in the substrate surface), the sites at which adatoms become immobilized are those that increase their atomic coordination, which leads to adatom incorporation at steps and to the formation of an island wherever two adatoms meet. Therefore, species deposited on top of an existing layer have a strong tendency for transferring to the next lower level by a mechanism known as interlayer mass transport [93]. Depending on the type of bonding, the probability of this transfer over a step-edge may be reduced by the energetically unfavorable position of an adatom at the step edge, leading to a so-called Ehrlich-Schwoebel (ES) barrier [94,95]. A very large ES barrier would immediately lead to three-dimensional (3D) growth, as voids (or holes) in each layer would only be filled at a rate of (1-1/$e$) per deposited monolayer. While it is conceptually easy to understand the formation of an ES-barrier for cases such as the growth of noble metals, there are arguments against such a barrier for more complex systems, such as compound semiconductors [96]. The frequently-observed two-dimensional growth habit of complex oxides, such as $SrTiO_3$, by a variety of techniques, argues against a behavior dominated by an ES-barrier for these materials.

A realistic surface will always exhibit step terraces spaced on average at $L \approx a/\alpha$, where $\alpha$ is the miscut angle of the substrate (in radians) and $a$ is the material's lattice parameter. In the case of low deposition flux $F$ and a high diffusion rate (i.e., for large values of $D/F$), adatoms will have a tendency to migrate to the step terraces without nucleating islands. The resulting step-flow is illustrated in Fig. 4. Adatoms can attach themselves to the steps on the left or the right of the terrace with probabilities $\kappa_L$ and $\kappa_R$, respectively. The steps $S_1$ and $S_2$ therefore travel with velocities $v_1 = \kappa_L F L_2 + \kappa_R F L_1$ and $v_2 = \kappa_L F L_3 + \kappa_R F L_2$. In the absence of a ES barrier, $\kappa_L = \kappa_R = \kappa$. In this case, step flow is unstable against perturbations: if we let $L_1 = L + \Delta L$, $L_2 = L - \Delta L$, and $L_3 = L$, then $v_1/v_2 = 2L / (2L - \Delta L) > 1$. In other words, the larger terrace grows faster than the narrower one, eventually leading to so-called step-bunching. In contrast, a large ES barrier (such that $\kappa_R \approx 0$) stabilizes step flow. It needs to be pointed out that in the case of heteroepitaxy, epitaxial strain also leads to step bunching [97], as has been confirmed by the careful comparison between experiments and calculations [98,99].

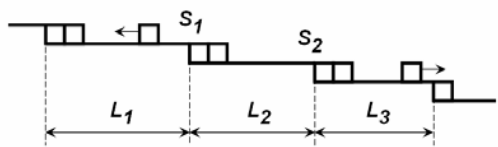

**Figure 4.** Step flow on a vicinal surface. Adatoms migrate either to the left ("up") or to the right ("down") to be incorporated at a step edge.

PLD differs from other physical vapor deposition methods such as MBE and sputtering in two important ways: first, the process is pulsed, meaning that a finite amount of material is deposited in a short time, namely the time of the plume interaction with the substrate. In contrast to the laser pulse, which lasts no longer than 30 ns even with excimer lasers, the plume interacts with the substrate for a few microseconds. Second, the energies of the impinging species in PLD are large and spread broadly with a typical mean energy of a few eV, while MBE provides a much more uniform energy distribution and energies of a few tens of an eV.

Some aspects of the energetic nature of PLD can be treated by considering an effective diffusivity $D'$ which may differ from the single-atom diffusivity and take into consideration collision-induced detachment from forming islands and other effects described below. The average deposition flux can be written as $F = N_p / \tau$, where $N_p$ is the amount of material deposited per pulse and $\tau$ is the time between pulses. With this notation, the requirement for step flow (namely that the time between laser pulses is large compared to the life time of diffusing atoms on a terrace) becomes $F < 2N_p D'/L^2$ [98]. In other words, whether or not step flow occurs depends not only on the diffusivity and the deposition rate, but also on the original miscut of the substrate.

For the growth of smooth, uniform layers, step-flow is clearly the preferred growth mode, as there are no issues regarding complete layer filling, nucleation of islands on top of existing islands, or even incorporation of defects at points of coalescence between islands. However, there are no periodic changes in surface characteristics that can be tracked in order to monitor *in situ* the growth and to terminate growth after completion of a predetermined number of layers. Therefore, for the synthesis of precise superlattice structures, ideal LBL growth is required. In light of this requirement, we now turn our attention to nucleation and growth of islands.

4.2 Nucleation and island growth

Acknowledging that epitaxial film growth derives virtually all of its utility from the possibility of heteroepitaxy, we nevertheless limit the following discussion to the special case of homoepitaxy. The homoepitaxial model system allows us to focus on pure kinetics in the formation of atomically sharp interfaces without interference from such issues as interfacial diffusion, strain relaxation by defect formation, and more than one material-specific surface free energy.

We start by considering the importance of the pulsed nature of the process. The question of energetics in the mechanism of PLD will be addressed later. In the simplest model, one assumes that whenever an adatom encounters another adatom as its nearest neighbor, both atoms stop diffusing and nucleate an island. The number of island grows quickly within the deposition of as little as a few percent of a monolayer, after which it changes insignificantly [100]. This is simply a consequence of the fact that the adatom density at the surface is reduced (at fixed flux) as the step edge and island densities increase. Simply put, the adatoms become more likely to encounter an existing island than another diffusing adatom after deposition of a few percent of a monolayer. For PLD, this implies that the number of nucleation sites saturates after the first laser pulse. Consequently, additional material will attach itself to the existing islands, the number of which essentially remains unchanged at least until coalescence of islands occurs. We will show below that ripening, i.e., the process by which larger islands grow at the expense of smaller ones is undesirable. In contrast, any mechanism that results in the formation of additional islands without increasing the nucleation density (i.e., breaking up of existing islands into smaller ones) would delay the formation of a second growth layer.

Figure 5 shows schematically the steps involved in the formation and growth of islands. Initial deposition (i.e., the first laser pulse) leads to the nucleation of islands at a characteristic spacing $\lambda$. Various mechanisms influence the value of $\lambda$, most importantly $D$ and $F$. Other factors may also play a role, such as the energetics of the impinging species or an additional energy source such as an added ion beam. For simplicity, we will only consider the case of islands that remain essentially immobile once they reach a critical size of a few atoms. An important point to remember when considering how to change the growth mode is that $\lambda < L$ must be satisfied to avoid step flow. Deposition of additional material then results in the growth of these islands, initially without nucleation of new islands neither on top nor between the existing ones in the ideal case of a negligibly small ES barrier. The island size remains below $\lambda$ until the point of coalescence. The very definition of $\lambda$ implies that new islands do not nucleate on an island with diameter $2R < \lambda$. Whether additional adatoms (or, in the case of perovskite, entire building blocks) formed on top of an island insert themselves by migration to the island edge or by lateral pushout of material is irrelevant at the earlier stages, where no second-layer nucleation is expected. At later stages, i.e., when the islands are relatively large, insertion anywhere other than the island edges will become energetically unfavorable. If deposition proceeds under unchanged conditions, new islands will invariably nucleate on top of the existing ones as soon as a critical island size is reached [101], i.e., when coalescence is reached at $R \approx \lambda/2$. This leads to the unavoidable formation of a two-level growth front. The ideal case of LBL growth, where one layer is completely filled before nucleation of new islands, is therefore not observed [101].

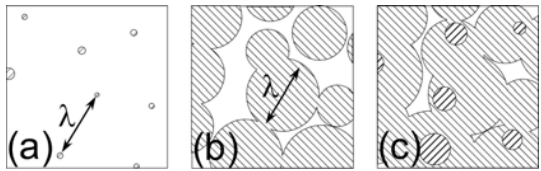

**Figure 5.** Top view of film growth in a simple view of nucleation, island growth, and coalescence. (a) Nucleation of islands with a characteristic spacing, $\lambda$. These islands form layer *n+1* on top of the previous surface (layer *n*). (b) Growth of islands leads to coalescence when the typical island size becomes comparable to $\lambda$. (c) Nucleation of the next layer (*n+2*) on top of the islands (layer *n+1*) begins when the island size is comparable or larger to the nucleation site spacing, $\lambda$.

Again, it is important to remember that we are concerned here with growth outside the step flow regime. Ideal LBL growth can thus not be achieved by increasing the adatom mobility (perhaps by using high-energy impinging species), which would promote step flow. It is nevertheless interesting to note that high cluster mobilities have been proposed as possible origin of enhanced growth kinetics in PLD [102].

The time-dependent distribution of adatoms on the growing surface is a key difference between MBE and PLD. In fact, while we would assume unchanged conditions during an ideal pulsed process such that the adatom concentration on top of the islands is identical to that between them, this is not the case for MBE: there, the edges of each terrace act as a drain, and in a steady-state process, the adatom concentration on an island is smaller than on an infinite surface [103]. This clearly aids in minimizing nucleation on top of existing islands. In a pulsed process, if the deposition occurs on a time scale that is shorter than that corresponding to the motion of the adatoms, no such self-limiting mechanism exists – the adatom density on top of an island is very close to that responsible for the initial nucleation at a characteristic distance, $\lambda$. We will return to this issue later when we show that the success of PLD is related to the fact that interlayer transfer (motion of material from the top of an island to the layer below) occurs largely at time scales comparable to the plume duration.

4.3 Ripening and island shape

It is clear from the above consideration that ripening leads to larger islands at the same surface coverage. It is undesirable in LBL growth because nucleation of the second layer becomes favorable at an earlier stage of growth [104]. Of course, some ripening is expected to occur in all processes where detachment of an adatom from an existing island is possible.

The picture of circular islands, as schematically drawn in Fig. 5, is a simplification that we have made without special justification. In fact, in the simple model mentioned above of adatoms "frozen in place" as soon as they encounter another adatom, the shape of islands is anything but circular: numerical simulations indicate the formation of dentritic growth patterns [100]. Such growth patterns would be advantageous, since for a given fractional coverage of the surface, the path from any point on top of an island to an available edge site is, on average, smaller than in the case of circular islands.

During the early part of a monolayer deposition, most of the island growth occurs via migration of adatoms on the initial surface towards the edges of the islands (rather than by interlayer transfer down from the second level). It is therefore easy to assume that the edges of each island indeed first assume a somewhat dentritic (or fractal) shape (assuming that the time scale for diffusion of an individual adatom before attachment to an edge is short as compared to that required for diffusing along a step edge). The island contours then smoothen during the time between two laser pulses. Such a smoothening process will reduce the density of step edges, and result in a recovery of the RHEED signal transients that occur at each laser pulse. Clear evidence of the formation of smooth contours comes from *ex situ* AFM observations of growth surfaces, which are often dominated by isotropic islands.

A thermally-driven process that reduces the step edge density for a given surface coverage clearly hinders ideal LBL growth. One way to avoid this mechanism is to reduce the time between laser pulses, which, for example, can be achieved in a method termed "pulsed laser interval deposition" [105]. Here, the amount corresponding to exactly one monolayer is rapidly deposited, followed by a pause during which some "annealing" of the surface is allowed. Deposition energetics may also play an important role. Aziz and coworkers have performed careful comparisons between MBE-grown and PLD-deposited metal and semiconductor films, using various PLD conditions [106,107]. As the pulsed nature of PLD by itself is insufficient to explain the observed behaviors, energetic effects are found to be important in determining the evolution of surface morphology. Similarly, the effect of energetic species in removing adatoms that were attached near the edges of a growing island has been considered [108]. A model in which impinging species break up existing small islands is found to be compatible with SXRD data in heteroepitaxial growth [109]. Similarly-obtained data have also been explained in terms of a non-thermal smoothing mechanism even in an experiment where growth evolves beyond a two level system after the deposition of a few monolayers [110]. The precise analysis of the growth front evolution, however, can only be achieved by quantitative measurements of layer coverages, to which we now turn our attention.

## 5. Monitoring of growth kinetics

5.1 Introduction and RHEED studies

In ideal superlattice growth it is not only assumed that the starting surface is prepared atomically flat, but switching from one materials to the next is performed *exactly* at the top of a RHEED oscillation—the supposition here is that the growing layer is complete. However, numerous examples show that this assumption is almost never fully justified. Here we use the results of recent growth kinetics studies to examine this issue in more detail.

We start this discussion by noting that the RHEED oscillation maxima do not correspond with layer completion [111]. This is not a serious problem when RHEED is used for counting layers or determining growth modes. However, as the number of layers becomes smaller this fact must be recognized and properly accounted for properly. Rigorous consideration of this subtle effect is particularly important in growth kinetics studies. RHEED oscillations (or intensity oscillations in other diffraction techniques) are associated with the periodic nature of the layer filling process that occurs when growth proceeds in a cyclic fashion. Such a cyclic process does not, however, necessarily imply ideal LBL growth, but can occur in the form of overlapping two-layer growth as described below (analogous to the two-layer behavior observed under sputter removal of material [112]). Only in perfect LBL growth do these oscillations have a maximum intensity equal to the starting intensity when the layer is full and minimum intensity when the layer is half filled.

Simple arguments concerning nucleation, growth, and coalescence, developed by Comsa and coworkers in connection with metal MBE, show why the maximum intensity in diffraction measurements never again reaches its starting value [10]. The intensity maxima occur at the point when more atoms remain in the top islands than are added to the base islands. The value of this maximum is always less than the starting intensity because even if the base layer is complete—which is almost never the case—there are already islands on top of the base layer that reduce the intensity. The situation becomes more complex if the base layer filling is slow and the requirement for nucleation of a new layer on top of the growing layer is satisfied before the base layer is complete. This interface broadening requires that the turning points in the intensity oscillations (minima and maxima) now be determined by adding up the contributions from all the open layers. Of course, the intensity envelope continues to decay as the interface broadens, with the extreme case being three-dimensional (3D) growth that eliminates all intensity oscillations.

If the supply of atoms is interrupted in LBL growth, the response of the surface in filling the open layers results in a recovery of the intensity toward the initial value. This recovery was observed first in GaAs, and was found to consist of two steps with distinctly different time-constants [113]. After considerable debate a consensus emerged that the initial rapid step corresponds to filling of holes by atoms transferring into the base layer, and the long time constant step corresponds to ripening of the remaining islands [114]. However, it must be noted that this simple picture is significantly altered in the presence of an ES barrier that impedes the transfer of atoms from the top layer into the base layer (see section 4.1 above).

It is naturally tempting to think of PLD as an opportunity to study the mechanisms associated with the recovery in oxide materials by using time-resolved measurements: in a simplistic (and, as we will show, inaccurate) view, deposition occurs on a fast time scale determined by the arrival of the species in the plume, and growth occurs after these species arrive on the surface, nucleate islands or diffuse around in search of the proper crystallographic lattice sites. Detailed measurements of the time constants would then provide important clues for the identification of the possible mechanisms involved. In early RHEED intensity oscillation measurements the amount of material deposited per pulse was too small to cause modulation (steps) in the RHEED oscillations. In contrast to GaAs, in PLD of $SrTiO_3$ [89] and $YBa_2Cu_3O_{7-x}$ (YBCO) [115] only a single-step recovery was observed after growth termination. This recovery was attributed to a reduction of the step edge density by surface rearrangements that results in the formation of larger islands.

True time-resolved measurements of the recovery after single laser shots became possible with the ability to increase the amount of material deposited per pulse. Using time-resolved RHEED intensity oscillation measurements in YBCO PLD, single shot recovery was observed but only in the last shot before the peak of the oscillation period (assumed to be near full coverage) [116]. The temperature dependence of the time constants (ranging from 0.2 to 0.4s) would allow attributing the recovery to crystallization of the deposited material, a process fundamentally different from the previous interpretation of step edge density reduction. Complicating the unambiguous identification of the responsible mechanism is the fact that observation of fast processes associated with the plume arrival by RHEED is difficult, primarily because the acquisition time is limited to 30 – 100 ms by the CCD frame capture rate. A first indication of a fast recovery process was observed during growth mode studies of $SrTiO_3$ that mapped a wide range of substrate temperatures and deposition conditions [117]. This work observed only partial recovery of the RHEED intensity, but systematically measured the slow recovery stage. Additional scanning tunneling microscopy data confirmed that the recovery was related to slow surface migration.

5.2 Growth Kinetics Studies by Surface X-ray Diffraction

RHEED is not an ideal tool for growth kinetics studies because the strong interaction of the electrons with the surface causes multiple scattering and requires dynamical theory for rigorous interpretation of the intensities [111,118]. In contrast, the above-mentioned intuitive and widely used step edge density model emerged from empirical correlation between STM imaging of step density and RHEED intensity [119,120]. As an alternative technique, SXRD has the unique advantage that kinematic scattering is applicable and that the intensity can be interpreted directly in terms of surface coverage [121,122]. Time-resolved SXRD at crystal truncation rod (CTR) positions allows real-time measurements of interface layer formation, so the combination of SXRD and PLD represents a powerful technique for gathering unique information on interface formation and growth kinetics [123-125]. X-rays also have unique practical advantages. The ability of x-rays to avoid surface charging makes x-ray diffraction the most suitable technique for studying oxide surfaces and interfaces [126]. Unlike electrons, x-rays are not scattered by the high pressure background that is often necessary in oxide growth and do not interact with the surface to chemically perturb or alter the growing film [127]. Finally, the static surface structure of $SrTiO_3$ has been studied previously by measuring crystal truncation rods [128,129] to provide background information on the state of the starting surface.

The scattering conditions in our SXRD experiments are set to monitor the formation of $SrTiO_3$ unit cells. The scattered intensity is measured simultaneously at the specular (0 0 ½) and off-specular (0 1 ½) CTRs before, during, and after PLD growth of $SrTiO_3$. The significance of measuring both rods is that the specular rod has momentum transfer along the surface normal and provides information only about deposition, i.e., the height distribution of material. The lateral ordering on the surface, and in-plane registry with the lattice that is synonymous with crystal growth is confirmed by measuring an off-specular rod (h,k) ≠ (0,0) which has an in-plane momentum transfer component. Well-developed and persistent RHEED-like SRXD growth oscillations are observed simultaneously at both specular and off-specular CTR positions during homoepitaxial growth of $SrTiO_3$ at temperatures ranging from $310^oC$ to $780^oC$ [130].

New experimental capabilities enable the measurement of SXRD transients with 10 μs time resolution [121]. The ability to measure the crystalline layer formation on the same time scale as the plume arrival time reveals new details that advance our understanding and change the traditional view on how PLD works. These fast measurements are made possible primarily by the high brilliance of a third-generation synchrotron x-ray source at the Advanced Photon Source [131]. A critical factor that enables taking full advantage of the high intensity is the perfection of the initial surface. We developed a highly selective screening process for choosing the substrates used in the SXRD growth experiments – yielding 1 usable substrate out of every 3 after annealing and AFM inspection. The specular rod intensity at the typical growth temperature ($620^oC$ to $650^oC$) is on the order of $10^6$ counts/s. The value of the initial SXRD signal is a good predictor of the film growth quality as judged by the persistence of the intensity oscillations after initial decay.

The high-resolution SXRD transients in Fig. 6 show that the fast stage in PLD cannot be resolved even with a microsecond range time resolution. These measurements indicate that crystallization and—as we explain below—a large fraction of the total observed interlayer transport occur on the time scale of the plume arrival, much faster than previously known from using RHEED and SXRD [110,130,132]. Thus, contrary to the simplistic picture mentioned above, deposition and growth (crystallization) cannot be separated by the pulsed nature of the PLD process [121]. Surface migration preceding crystallization would manifest itself as a temperature dependent time delay between the fast steps in the specular and the off-specular transients [133].

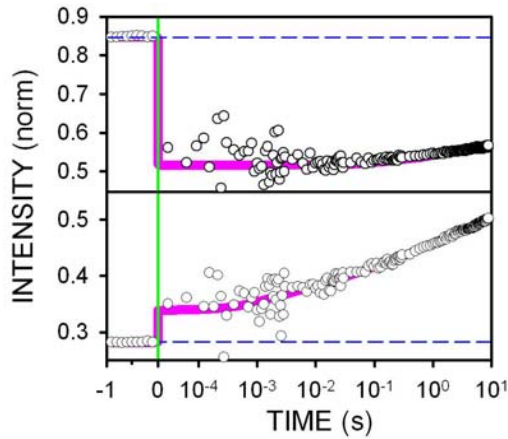

**Figure 6.** Single laser shot time-resolved SXRD transients at the (0 0 ½) specular rod. The top curve showing a drop corresponds to a laser shot following the growth intensity oscillation maximum, while the bottom curve showing a jump corresponds to a laser shot following a growth intensity oscillation minimum. Each curve represents an average of 10 laser shots. The full oscillations are shown in Fig. 8. A logarithmic time scale, which spanned from 1s before the laser shot to 10s after, was used to capture the SXRD transient. As a point of reference, the transients immediately after the laser shot were captured using a sampling time of 6μs, and the data are shown binned to 25μs. Note that because of the logarithmic time sampling the statistical fluctuations of the data appear exaggerated at short sampling times. The dashed lines correspond to the x-ray intensity before the laser shot and the vertical solid line marks the time when the laser was fired. Reprinted with permission from Ref. 121, Copyright 2006, American Physical Society.

5.3 Quantitative Measurements of Time-Dependent Coverages

In previous work both RHEED and SXRD growth intensity oscillations were analyzed using transport models [110,130]. On qualitative level these models provide a clear illustration of the importance of interlayer transport, and give invaluable clues for understanding the characteristic features of the SXRD transients in terms of interlayer transport [111]. We consider these models here despite the number of examples showing that they do not fit the experimental data. It can be shown in simple terms that for a two level system the diffraction signal is most sensitive to interlayer transport near full coverage [111]. The rate of interlayer transport in the recovery is given by $d\theta_2/dt = k\theta_2(1-\theta_1)$, where $\theta_{1,2}$ are the coverages in layers 1 and 2, respectively, and k is the interlayer transport rate constant. This differential equation can be solved to determine the recovery for each particular value of the initial coverage $\theta_1$. Approaching full coverage, the $(1-\theta_1)$ term dominates the recovery because fewer holes remain in the base layer that can be filled. The important conclusion from the analysis of the simplest possible model is that the appearance of the recovery signal is affected not just by the time constant but by the coverage at which the recovery is observed.

The key advantage of SXRD is that it enables quantitative determination of the coverages directly from the measured intensity without specific assumptions about the physics of the growth process such as, for example, the shape of the islands. The use of the kinematic approximation enables straightforward calculation of the scattered intensity within certain constraints that are determined only by the number of incomplete layers during growth. It can be shown that for one or two layers on top of the substrate—islands on top of a substrate, or islands on top of a base layer with holes—the intensity change from material distributed between these layers can be calculated exactly [121]. For more than two layers there are an arbitrary number of possibilities. In contrast, RHEED intensities must be calculated in terms of step edge density [111,134]. This calculation requires assumptions about the shape of the islands to account for how the step-edge density changes with coverage. Note that the specularly scattered x-ray intensity depends only on the number of scatterers and is independent of the shape of the islands.

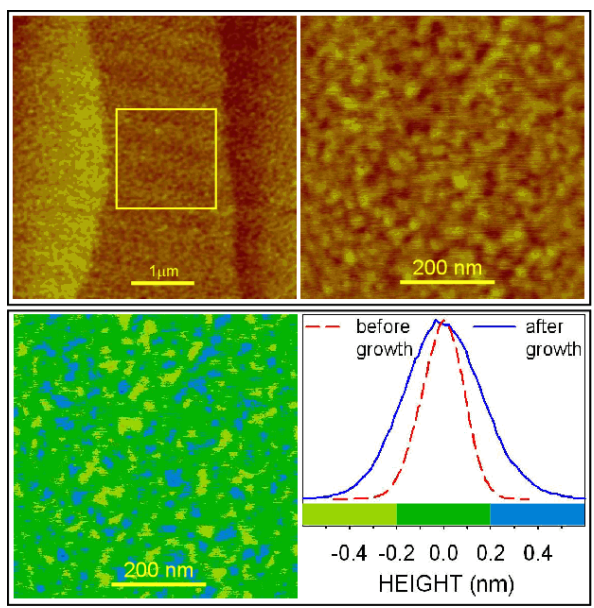

**Figure 7.** AFM image that illustrates the range of surface morphologies observed in the three-level growth mode. To eliminate the possible influence of terrace steps when leveling the raw AFM image, the scans were performed on a single terrace as designated by the yellow square. The dashed line in the histogram shows the surface width of the substrate, and the solid line that of the film. A color scheme shown in the histogram was used to convey what the x-rays are "seeing" at the anti-Bragg condition by setting solid color changes to occur at one unit cell heights. The image was obtained from a film that was more than 100 unit cells thick, and grown with a 50s dwell time.

The validity of the assumptions for calculating the coverage from the SXRD intensity, namely that no more than three levels need to be considered, was confirmed by systematic AFM imaging of the substrates before and immediately after film growth. Fig. 7 shows an AFM image that illustrates the quality of the film surface observed after film growth. In addition to showing the raw image, which at first appears to show only random noise on the data obtained within a single terrace, we use a special coloring scheme in which each solid color represents a one unit cell step to convey "what the x-rays are seeing." The interface width is also plotted as a height histogram. The FWHM of the starting surface is typically around 0.2 nm. Interface roughness shows up as broadening of the histogram and the appearance of tails that extend past 0.2 nm on both sides indicating the presence of holes and islands on the surface. This clearly illustrates that the data is consistent with a three-level (two-layer) model and excludes the possibility of roughening beyond these two layers. The interface broadening after film growth was compared with the starting surface for the same substrate for numerous growth runs. There is a clear correlation between the persistence of the intensity oscillations and growth front broadening observed by AFM imaging. Samples that exhibit persistent intensity oscillations show minimal or no measurable growth front broadening compared to the substrate, and many samples show surfaces that are significantly smoother than that shown in Fig. 7, occasionally exhibiting a narrower height histogram of the starting surface.

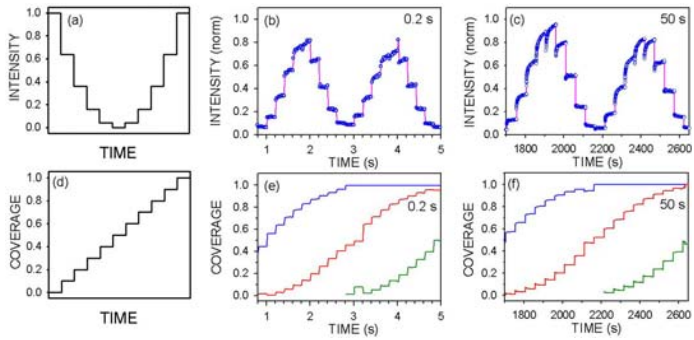

**Figure 8.** Ideal layer filling in pulsed mode is compared with actual PLD growth of STO at 0.2s and 50s dwell time. The intensity oscillations are shown in the top row, and the time-dependent layer coverages are given in the bottom row. Note the apparent similarity between one full period of actual growth (center part of (b) and (c)) with ideal LBL growth shown in (a). Interlayer transport is manifested in the subtle features in which the real data departs from ideal behavior. The time dependent coverages extracted from the single shot transients are given in (e) and (f). Along a full set of layer filling data in the middle, layer filling near completion is given in the top curve and the onset of growth is shown by the bottom curve. Compare the ideal staircase in (d) with the actual data in (e) and (f) to see the curvature change near layer filling and at the onset of growth. Adapted from Ref. 121.

Instead of using a transport model to fit the data, the time-dependent coverages are calculated directly from the SXRD intensities [121]. At each point, the fractional coverage $\theta_n(t) = (\sqrt{I(t)} + 1 + 2\eta(t))/4$ of layer $n$ is calculated directly from the normalized intensity $I(t)$ and the deposited amount of material $\eta(t) = \theta_n(t) + \theta_{n+1}(t)$, which increases with each laser pulse by a fixed step height. Without making any further assumptions, this method will properly distinguish between true layer-by-layer growth and a three-level system.

The corresponding time-dependent coverages shown in Fig. 8 resemble a rising staircase with each step having a unique shape. The shape within each staircase step indicates whether net interlayer transport occurs into or out of the layer and roughly falls into three categories. An ideally flat shape indicates no net interlayer transport and occurs near 0.5 coverage. A slightly upward curving step indicates interlayer transport into the layer and occurs above 0.5 coverage. A slightly downward curving step indicates interlayer transport out of the layer and occurs below 0.5 coverage. The curvature of the non-flat steps becomes more pronounced with increasing dwell time between successive laser shots, serving as a qualitative indicator of the rate of interlayer transport.

In perfect LBL growth, coverage of layer $n$ would reach 100% before layer $n+1$ nucleates, which is clearly not observed. Nevertheless, by the time $\theta_{n+1}$ reaches a value of, for example, 0.3, the layer below is more than 90% complete ($\theta_n > 0.9$): the surface at this point consists primarily of islands on a base layer having a few holes. Nucleation of the next layer always begins before completion of the previous one, which means that a truly atomically flat growth surface is never observed. Further evidence of this behavior is obtained from the analysis of diffuse x-ray scattering [135], which shows a gradual increase in the characteristic length scales on the surface: the system indeed maintains a "history," which would not be the case in homoepitaxy if perfect surfaces were formed.

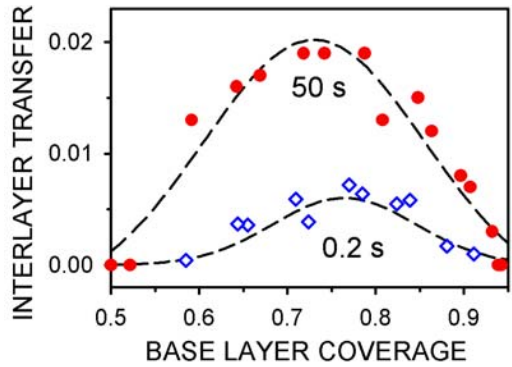

**Figure 9.** A comparison of thermally driven interlayer transport at a fixed temperature and dwell times of 50s (solid dots) and 0.2s (diamonds) plotted as a function of the coverage in the base layer into which the transport occurred. The deposition rate is 0.1 unit cell (uc) per pulse. The dashed Gaussian lines are guides to the eye. The peak corresponds to the maximum amount of thermally transferred material, which is 0.02(uc)/0.1(uc) = 20% of a single shot at 50s dwell time, and less than 5% at 0.2s dwell time. Adapted from Ref. 121.

The significance of the single shot time-dependent coverages is that they can be used to determine for each laser shot the amount of material that is transferred during the dwell time between laser shots. These coverage changes can be used for constructing a picture about the role of the thermally-driven slow interlayer transport process in $SrTiO_3$ PLD, i.e., the mechanisms that occur within the experimentally resolved time scales above 25 μs. This thermally-driven interlayer transfer occurs in addition to a fast (and experimentally inaccessible) component. The plot in Fig. 9 shows the amount (in terms of coverage) of material transferred by the slow process from the top of the islands into the base layer as a function of coverage in the base (growing) layer. Data is shown for dwell times of 0.2s and 50s between consecutive laser pulses. Below half-coverage, no material remains on top of an island (otherwise, strong roughening would occur immediately and our three-level model would fail), implying complete interlayer transfer into the layer below the island. However, just above half-coverage, the slow interlayer transfer is still negligibly small, not reaching its maximum value until a coverage of about 0.7 is reached. The value at the maximum of the curves gives the largest fraction of the coverage that was transferred by this slow process. For example, at the peak in the data obtained with a 50s dwell time, a coverage change of 0.02 can be attributed to this slow process. One a per shot basis, this corresponds to 20% in the present case since there are 10 shots per monolayer. For a dwell time of 0.2s, however, this value falls to 5% for a 0.2s dwell time [121]. This is a singularly important trend for practical film growth because it shows that at typical PLD repetition rates of few Hz the fraction of thermally driven interlayer transport is small (albeit not entirely negligible). Most importantly, there is no significant distinction between the coverage results obtained for different dwell times, which means that the thermally-driven interlayer transfer is not the dominant mechanism leading to smooth film growth, nor is it a necessary component. Similar behavior with an absence of recovery was observed in Ge PLD [136], prompting an interpretation that non-equilibrium laser driven processes occurred on the same or a shorter time scale than the arrival of the laser plume. Remembering that a study of the fast step discussed above already established that crystallization in $SrTiO_3$ occurs on a time scale of microseconds it is clear that the fast interlayer transport component must be the basis of any growth kinetic manipulation method aimed at obtaining atomically sharp interfaces. Another argument in favor of this view is that the thermally-driven interlayer transport component is a difficult to control slow process, and most importantly, its effects on interface broadening are largely unpredictable because the phenomenon is not well understood.

5.4 Kinetic Growth Manipulation

In this section we describe kinetic growth manipulation as a method for obtaining atomically sharp interfaces. Simply stated, the goal of kinetic growth manipulation is to achieve near-perfect LBL growth by delaying the nucleation of islands on the top of the growing layer for as long as possible (ideally until the growing layer is complete) [137]. The PLD process has been studied extensively and various approaches have been explored to reach this goal, which would manifest itself by complete recovery of the RHEED intensity after each unit cell deposition. Even in the more elaborate techniques, such as the previously-mentioned "interval deposition" approach [105], this has not yet been achieved.

Our discussion of the time-resolved SXRD data illustrate two features that are contradictory to conventional wisdom regarding PLD: First, crystallization occurs in the first few microseconds after arrival of the deposited material, i.e., *during* the plume/substrate interaction (of course, some processes leading to a low defect density will still occur *between* laser pulses, but they are more akin to sintering mechanisms than to crystallization). Second, the most important component of interlayer transfer also occurs during the plume/substrate interaction. The slower (thermal) mechanisms of interlayer transfer, which also lead to ripening and island growth, are therefore not only undesirable, but also unnecessary.

The unique advantage of PLD as a tool for kinetic growth manipulation comes from the extremely large dynamic range of the instantaneous growth rates [5]. By simply adjusting the laser parameters, the growth rates in PLD can be varied over several orders of magnitude. As the island density and the characteristic nucleation length scale change inversely with the deposition rate, this ability to vary the characteristic nucleation length scale enables a different scheme for growth kinetic manipulation: the interlayer transport can be controlled in the critical stages of the layer filling process according to the growth kinetic picture discussed above.

Methods to manipulate growth by modulating the nucleation density have been discussed previously for the case of metal epitaxy [138]. The power law dependence of island density on deposition flux renders the method rather inefficient in many techniques. In PLD, however, the growth rate is readily modulated by orders of magnitude simply by varying the parameters of the laser spot on the target. Therefore, the first step in an efficient growth manipulation scheme is to induce nucleation of a high density of islands with the first laser shot. This creates a large number of small islands that are close to each other. The tendency of islands to ripen during the time $t_c$ from island nucleation to coalescence limits the optimal island density to $\sim 1/\lambda^2 \cong 1/D_s t_c$, where $D_s$ represent the surface diffusivity. As $D_s$ depends on temperature, even in the highly nonequilibrium PLD growth technique, the nucleation length scale depends on the substrate temperature. However, the fact that both interlayer transport and crystallization occur on a microsecond time scale (i.e., orders of magnitude faster than $t_c$) allows us to suppress the thermally-driven processes simply by increasing the growth rate.

Therefore, once the island density is set by the first laser pulse, the thermal processes that would lead to ripening must be minimized by rapid layer filling. However, the deposition per pulse must be maintained such that the critical nucleation length is always larger than the island size. As we have discussed, nucleation of new islands on top of the growing islands cannot be totally avoided after coalescence, but it can be substantially reduced by maximizing the characteristic nucleation length scale. Realization of this step requires a substantially reduced amount of material deposited by each laser pulse necessary to complete the layer. Note the important distinction between total growth rate (which must be high such as to avoid ripening) and growth rate per pulse (which must be low such as to minimize nucleation of new islands).

Therefore, the observation that "pure" PLD is essentially a process in which deposition and crystallization occur simultaneously (with some undesired thermal processes taking place between laser pulses) allows us to postulate the following recipe for optimized growth: the nucleation density can be set during a first pulse (with parameters chosen such as to maximize the amount of material deposited during the first pulse). The remainder of the layer is then grown under conditions where each laser pulse deposits a significantly reduced amount of material, while the deposition rate is kept high by increasing the laser repetition rate.

## 6. Conclusions

Pulsed laser deposition has been tremendously successful in the synthesis of complex-oxide materials. As we have shown in this review, the process is easily adapted to a broad range of specialized methods, including the formation of alloys, compositional-spreads, and superlattices. The analysis of physical properties arising at interfaces (and thus in superlattices) is often based on the assumption of atomically-flat junctions between dissimilar materials. Consistent with earlier studies on epitaxial growth, our analysis shows that the required perfect layer-by-layer growth is never achieved. However, careful analysis of our time-resolved SXRD data demonstrates an important property of PLD that renders this method particularly suitable for obtaining abrupt interfaces: in PLD, crystallization and a majority of interlayer mass transport occur during the deposition of the material within the arrival time of a single-pulse plume and not—as one might be tempted to assume—during the dwell time between successive laser pulses. By studying the kinetics of interlayer transport we identify a regime where nucleation and growth of thin films is driven by nonequilibrium processes during the time of plume/substrate interaction, with no thermal contributions provided by substrate heating. This encourages us to develop kinetic growth manipulation schemes based on varying the amount of material deposited with each pulse depending on the degree of layer filling at each time. Our results also shows how—in conventional PLD—these non-thermal processes successfully avoid roughening beyond the three-layer growth front, due to the independence of interlayer transport on the thermal processes responsible for island size ripening.

Acknowledgements:
The authors would like to acknowledge stimulating discussions and important contributions by our collaborators M.D. Biegalski, Wei Hong, D.H. Kim, B.C. Larson, H.N. Lee, D.H. Lowndes, I. Ohkubo, C.M. Rouleau, Zhigang Suo, J.Z. Tischler, M. Yoon, Z. Zhang, and P. Zschack. This work was supported by the Division of Materials Sciences and Engineering, Basic Energy Sciences, U.S. Department of Energy, under contract DE-AC05-00OR22725. The use of the APS was supported by the U.S. Department of Energy, Office of Science, Office of Basic Energy Sciences, under Contract No. DE-AC02-06CH11357.## References

[1] Rao C N R and Raveau B 1998 *Transition Metal Oxides 2nd Edition* (New York: Wiley-VCH)

[2] Chrisey D B and Hubler G K ed 1994 *Pulsed Laser Deposition of Thin Films* (New York: John Wiley & Sons)

[3] Lowndes D H, Geohegan D B, Puretzky A A, Norton D P and Rouleau C M 1996 *Science* **273** 898

[4] Willmott P R and Huber J R 2000 *Rev. Mod. Phys.* **72** 315

[5] Willmott P R 2004 *Prog. Surf. Sci.* **76** 163

[6] Lee H N, Christen H M, Chisholm M F, Rouleau C M and Lowndes D H *Nature* **433** 395

[7] Yamada H, Kawasaki M, Ogawa Y and Tokura Y 2002 *Appl. Phys. Lett*. **81** 4793

[8] Schlom D G, Haeni J H, Lettieri J, Theis C D, Tian W, Jiang J C and Pan X Q 2001 *Mater. Sci. & Eng*. B **87** 282

[9] Warusawithana M P, Colla E V, Eckstein J N and Weissman M B 2003 *Phys. Rev. Lett.* **90** 1586

[10] Rosenfeld G, Poelsema B and Comsa G 1997 Epitaxial growth modes far from equilibrium G*rowth and properties of ultrathin epitaxial layers, the chemical physics of solid surfaces* vol 8 ed D A King and D P Woodruff (Amsterdam: Elsevier) p 66

[11] Norton D P 2004 *Mater. Sci. & Eng. R* **43** 139

[12] Smith H M and Turner A F 1965 *Appl. Opt.* **4** 147

[13] Schwarz H and Tourtellotte H A 1969 *J. Vac. Sci. Technol.* **6** 373

[14] Desserre J and Floy J F 1975 *Thin Solid Films* **29** 29

[15] Zaitsev-Zotov S V, Martynyuk R A and Protasov E A 1983 *Sov. Phys. Solid State* **25** 100

[16] Dijkkamp D, Venkatesan T, Wu X D, Shaheen S A, Jisrawi N, Min-Lee Y H, McLean W L and Croft M 1987 *Appl. Phys. Lett.* **51** 619

[17] Sankur H, Gunning W J, DeNatale J and Flintoff J F 1989 *J. Appl. Phys.* **65** 2475

[18] Ohkubo I, Matsumoto Y, Hasegawa T, Ueno K, Itaka K, Ahmet P, Chikyow T, Kawasaki M and Koinuma H 2001 *Japan. J. Appl. Phys. Part 2 (Lett.)* **40** L1343

[19] Willmott P R, Manoravi P and Holliday K 2000 *Appl. Phys.* A **70** 425

[20] Koinuma H, Kawasaki M, Ohashi S, Lippmaa M, Nakagawa N, Iwasaki M, and Qiu X G 1998 *Proc. SPIE* **3481** 153

[21] Tanaka H and Kawai T 2000 *Appl. Phys. Lett.* **76** 3618

[22] Ohashi S, Lippmaa M, Nakagawa N, Nasagawa H, Koinuma H and Kawasaki M 1999 *Rev. Sci. Instrum.* **70** 178

[23] Chen P, Xu S Y, Lin J, Ong C K and Cui D F 1999 *Appl. Surf. Sci.* **137** 98

[24] Yang G Z, Lu H B, Chen F, Zhao T and Chen Z H 2001 *J. Cryst. Growth* **929** 227

[25] Holzapfel B, Roas B, Schultz L, Bauer P and Saemann-Ischenko, G 1992 *Appl. Phys. Lett.* **61** (26) 3178

[26] Égerházi L, Geretovszky Zs and Szörényi T 2005 *Appl. Surf. Sci.* **247** 182

[27] Kelly R and Miotello A 1994 *Pulsed Laser Deposition of Thin Films*

[28] Miotello A and Kelly R 1999 *Appl. Phys.* A **69** S67

[29] Geohegan D B 1994 Diagnostics and characteristics of laser-produced plasmas *Pulsed Laser Deposition of Thin Films* ed D H Chrisey and G K Hubler (New York: John Wiley & Sons, Inc.)

[30] Puretzky A A, Geohegan D B, Jellison Jr K G E and McGibbon M M 1996 *Appl. Surf. Sci.* **859** 96

[31] Knauss L A, Christen H M and Harshavardhan K S *unpublished*
23